\documentclass[aps,superscriptaddress,twocolumn,showpacs,UTF8]{revtex4-1}

\usepackage{graphicx}
\usepackage{dcolumn}
\usepackage{bm}
\usepackage{upgreek}
\begin{document}
\title{Sympathetic cooling of $^{40}\textbf{Ca}^+$ - $^{27}\textbf{Al}^+$ ion pair crystal in a linear Paul trap}

\author{Jun-Juan Shang}
\author{Kai-Feng Cui}
\author{Jian Cao}
\author{Shao-Mao Wang}
\author{Si-Jia Chao}

\affiliation{
	Key Laboratory of Atom Frequency Standards,\\
	Wuhan Institute of Physics and Mathematics,\\
	Chinese Academy of Sciences, Wuhan 430071, P. R. China 
}
\affiliation{
	State Key Laboratory of Magnetic Resonance and Atomic and Molecular Physics,\\
	Wuhan Institute of Physics and Mathematics,\\
	Chinese Academy of Sciences, Wuhan 430071, P. R. China 
}

\author{Hua-Lin Shu}
	\email{shl@wipm.ac.cn}
\author{Xue-Ren Huang}
	\email{hxueren@wipm.ac.cn}
\affiliation{
	Key Laboratory of Atom Frequency Standards,\\
	Wuhan Institute of Physics and Mathematics,\\
	Chinese Academy of Sciences, Wuhan 430071, P. R. China 
}
\affiliation{
	State Key Laboratory of Magnetic Resonance and Atomic and Molecular Physics,\\
	Wuhan Institute of Physics and Mathematics,\\
	Chinese Academy of Sciences, Wuhan 430071, P. R. China 
}
\affiliation{
	University of Chinese Academy of Sciences, Beijing 100049, P. R. China
}
\begin{abstract}
	The $^{27}$Al$^+$ ion optical clock is one of the most attractive optical clocks due to its own advantages, such as low blackbody radiation shift at room temperature and insensitive to the magnetic drift. However, it cannot be laser-cooled directly in the absence of 167 nm laser to date. This problem can be solved by sympathetic cooling. In this work, a linear Paul trap is used to trap both $^{40}$Ca$^{+}$ and  $^{27}$Al$^+$ ions simultaneously, and a single Doppler-cooled $^{40}$Ca$^+$ ion is employed to sympathetically cool a single $^{27}$Al$^+$ ion. Thus a "bright-dark" two-ion crystal has been successfully synthesized. The temperature of the crystal has been estimated to be about 7 mK by measuring the ratio of carrier and sideband spectral intensities. Finally, the dark ion is proved to be an $^{27}$Al$^+$ ion by precise measuring of the ion crystal`s secular motion frequency, which means that it is a great step for our $^{27}$Al$^+$ quantum logic clock.	
\end{abstract}

\maketitle
 
Single ion trapped in a Paul trap \cite{1,2,3,4,5,6} and neutral atoms trapped in an optical lattice \cite{7,8} are the two primary approaches for optical clock experiments. Optical clocks based on $^{27}$Al$^+$ [1], $^{171}$Yb$^+$ ion \cite{6} and $^{87}$Sr\cite{7} atom have possessed a fractional uncertainty of ${10}^{-18}$. The $^{27}$Al$^+$ ion clock based on the $^1S_0-^3P_0$ transition shows its own advantages \cite{9,10}. Firstly, the natural linewidth of $^3P_0$ state is only 8 mHz and the clock transition frequency is up to 1121 THz \cite{11}. Secondly, this transition possesses the fractional blackbody radiation (BBR) shift as low as $8(3) \times 10^{-18}$ at room temperature \cite{12}. Furthermore, both the  $^1S_0$ and $^3P_0$ energy levels of the $^{27}$Al$^+$ ion are insensitive to the magnetic drift \cite{13}. To date, the $^{27}$Al$^+$ ion cannot be Doppler cooled directly in the absence of the 167 nm laser. But this problem can be solved by sympathetic cooling with another ion species.

Sympathetic cooling is that one ion species is cooled by interacting with another ion which could be directly cooled. It has been widely studied experimentally \cite{14,15} and theoretically \cite{16,17}, including $^9$Be$^+-^{24}$Mg$^+$ \cite{18}, $^{40}$Ca$^+-^{115}$In$^+$ \cite{15}, $^{25}$Mg$^+-^{27}$Al$^+$ \cite{1} and $^{40}$Ca$^+$-$^3$Li$^+$ \cite{19}. All lasers applied to $^{40}$Ca$^+$ ion could be easily acquired from external cavity diode lasers. In addition, $^{40}$Ca$^+-^{27}$Al$^+$ quantum logic clock  has potential advantages in terms of second-order Doppler shift for ion traps with low heating rates \cite{17}.

In this Letter, we report recent progress on the sympathetic cooling of one $^{27}$Al$^+$ ion by one $^{40}$Ca$^+$ ion in a linear Paul trap. Both of them are created by two-photon ionization of neutral atoms. One $^{40}$Ca$^+$ and one dark ion are loaded to synthesize ``bright-dark'' ion pair cryatal. Using the ``four points locking scheme'' \cite{20} on the red sideband (RSB) and blue sideband (BSB), the mass of dark ion is calculated from the axial center of mode (COM) frequencies and it is proved to be $^{27}$Al$^+$ ion. The sideband-to-carrier intensity ratio of axial COM is 0.5, corresponding to the temperature of $^{40}$Ca$^+$- $^{27}$Al$^+$ ion crystal about 7 mK.

In a linear Paul trap, the confinement of charged particles is realized by an rf field in radial plane and one dc field along the $z$-axis. The total potential is given by\cite{21}:	
\begin{equation}
	\Phi(x,y,z,t)=\frac{U_0}{L^2}(z^2-\frac{x^2+y^2}{2})+\frac{V_0}{2}(1+\frac{x^2+y^2}{r_0})\cos(\Omega_{\rm RF})
\end{equation}	
where $U_0$ and $V_0$ are the DC and RF voltages respectively. 2$L$ is the distance between two endcap electrodes, 2$r_0$ is the distance between two diagonal bladelike electrodes, and $\Omega_{\rm RF}$ is the angular frequency of the RF field. When a single ion with a mass of $m$ and a charge of $Q$ is trapped, the axial and radial frequencies of the ion secular motion are given by	
\begin{equation}
	\omega_z=\sqrt{\frac{2 Q U_0}{m L^2}}
\end{equation}	
\begin{equation}
	\omega_r=\sqrt{\frac{Q^2 V_0}{2 m^2 r_0^4 \Omega _{\rm RF}^2}-\frac{1}{2}{\omega _z}^2}
\end{equation}	
If two ions are trapped simultaneously in a linear Paul trap, they will experience both the trapping potential and Coulomb repulsion potential. Then ions are Coulomb collided with each other successively, which leads to momentum transformation from one ion to another ion \cite{22}. The two ions will crystallize at their equilibrium positions along trap axis when they are cooled to the mK level. The remaining motion of two ions can be described as small coupled oscillations around their equilibrium positions. Along the trap axis, the motion consists of a superposition of breath mode (BM) and center of mode (COM). According to the approach described in Ref. \cite{17}, the oscillations are given by
\begin{equation}
	q_1 (t) = z_{\rm i}b _ 1\sin (\omega _{\rm i} t + \phi _{\rm i}) + z_{\rm o}b _ 2\cos(\omega _{\rm o} t + \phi _{\rm o} )
\end{equation}
\begin{equation}
	q_2 (t) = \frac{z_{\rm i}b _ 2}{\mu}\sin (\omega _{\rm i} t + \phi _{\rm i}) -\frac{z_{\rm o}b _ 1}{\mu}\cos(\omega _{\rm o} t + \phi _{\rm o} )
\end{equation}
where $\omega_{\rm i,o}$,$\phi_{\rm i,o}$ are the angular engenfrequencies and the phases of the COM and BM, and $b_1$, $b_2$ are the components of the normalized eigenvector of the COM and BM. $\mu$ = $\frac{m_2}{m_1}$ and $z_{\rm i}$, $z_{\rm o}$ are the mode amplitudes. Ignoring higher order nonlinear couplings, the mode frequencies along trap axis can be given by\cite{23}
\begin{equation}
	\omega _{z,\rm i}=\sqrt{\frac{\mu +1-\sqrt{\mu ^2-\mu +1}}{\mu }}
\end{equation}
\begin{equation}
	\omega _{z,\rm o}=\sqrt{\frac{\mu +1+\sqrt{\mu ^2-\mu +1}}{\mu }}
\end{equation}
\begin{figure}
	\begin{center}
		\resizebox{0.5\textwidth}{!}{\includegraphics{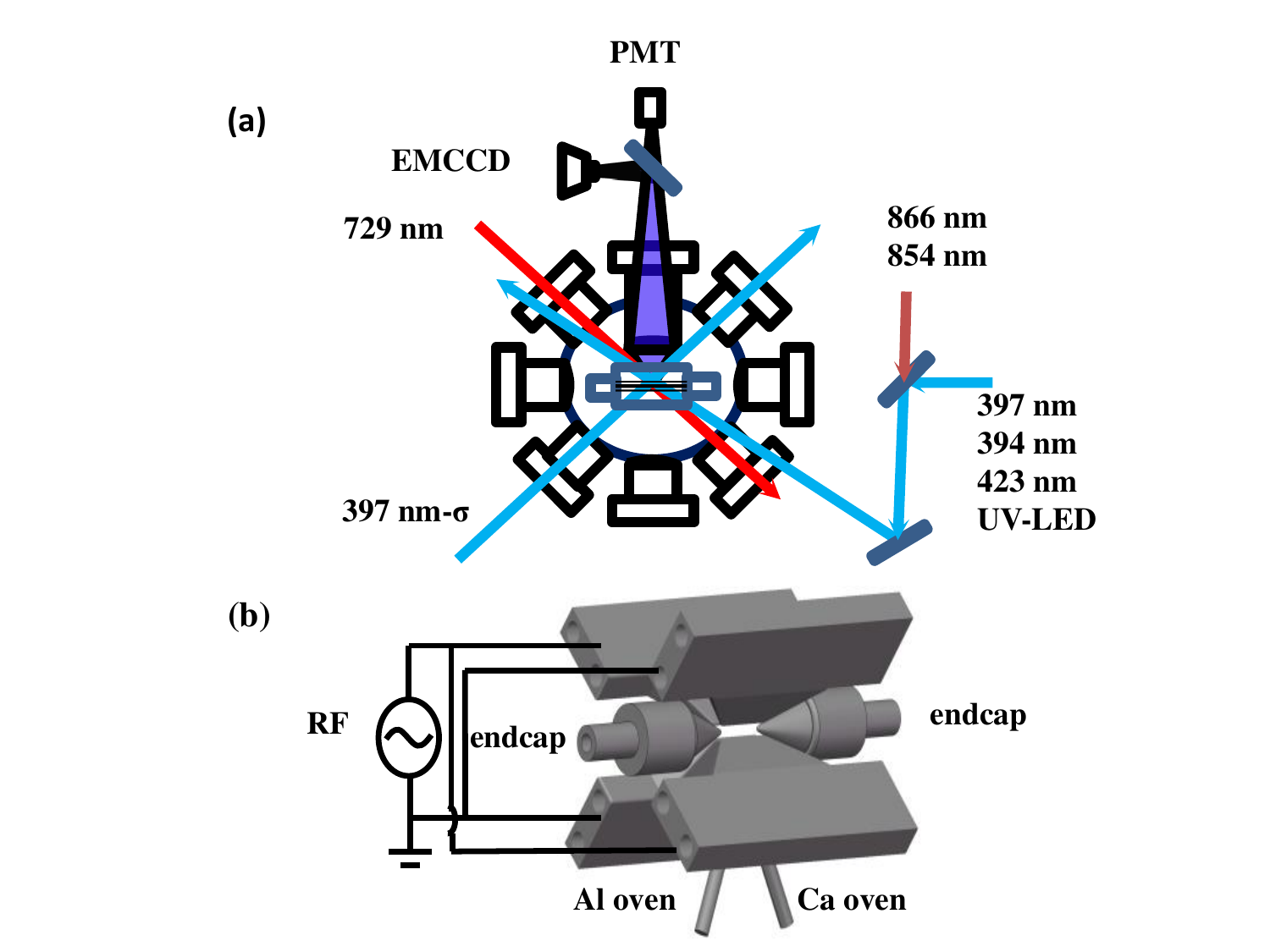}}
	\end{center}
	\caption{
		Top view of experimental setup and ion trap. (a) The direction of 729 nm laser is perpendicular to the bias magnetic and the 397-$\rm {\sigma}$ for state preparation is parallel to the bias magnetic. (b) The ion trap composes four blade-like electrodes and two endcap electrodes. EMCCD: electron multiplying charge coupled device; PMT: photomultiplier tube; and UV-LED: ultraviolet light-emitting diode.
		}
\end{figure}
The experimental setup is shown in Fig.1 (a). An UHV vacuum chamber is kept at $3\times 10^{-8}$ Pa by a 75 L/s ion pump. It consists a CF150 flange that acts as a base plate onto which an octagon is mounted. Six CF35 flanges are used as the viewports for the laser beams, and another two CF63 flanges are used as viewports for collecting the photons scattered by the ions. Ion fluorescence are detected by a photomultiplier tube (PMT) and an electron multiplying charge coupled device (EMCCD). As shown in Fig.1 (b), the linear Paul trap consists of four bladelike electrodes and two endcap electrodes made of stainless steel, and is characterized by $2 r_0$=1.6 mm and $2 L$=4.4 mm (Fig.1 (b)). On the top of the vacuum chamber, two CF35 feedthroughs are used to connect the RF driver with $V_{\rm PP}=$ 2000 V at 17.128 MHz and DC voltage of 150 V respectively. The secular frequencies $\omega_x$, $\omega_y$, $\omega_z$ of the trap for single $^{40}$Ca$^+$ ion are measured to be 2 MHz, 2 MHz and 300 kHz respectively. A stainless steel tube with 1.5 mm diameter and 0.1 mm thickness is used as the Ca oven. In contrast, an Al oven is more complicated, which is a thin alumina tube and twisted by a 0.3 mm diameter tungsten wire. Both the thin alumina tube and the tungsten wire are embedded into a larger alumina tube.

Both $^{40}$Ca$^+$ and $^{27}$Al$^+$ ions are created by two-photon ionization process after the Ca and Al atom ovens heated. When a $^{40}$Ca$^+$ ion is Doppler cooled by lasers at 397 nm and 866 nm, fluorescence at 397 nm is detected by a PMT and an EMCCD. The process of loading an $^{27}$Al$^+$ ion is similar to that of loading a $^{40}$Ca$^+$ ion. Laser at 394 nm is used to excite the Al atom from the state $^2P_{1/2}$ to $^2S_{1/2}$. Then another 394 nm photon is sufficient to reach the ionization threshold at 6 eV.

The number of dark ions can be confirmed by the distance between two adjacent ions, which depends on the charges of the ion and endcap voltage but not ion species\cite{24}. In the condition of 30 V endcap voltage, the relationship between ion distance and ion number was measured by trapping different numbers of $^{40}$Ca$^+$ ion shown in Fig.2(a). The ion's distance will increase as the number of ion reduces. In addition, it is helpful to verify the dark ion number that ions may exchange their positions along trap axis. Three $^{40}$Ca$^+$ ions and a dark ion are imaged with EMCCD, as shown in Fig.2(b).
\begin{figure}
	\begin{center}
		\resizebox{0.5\textwidth}{!}{\includegraphics{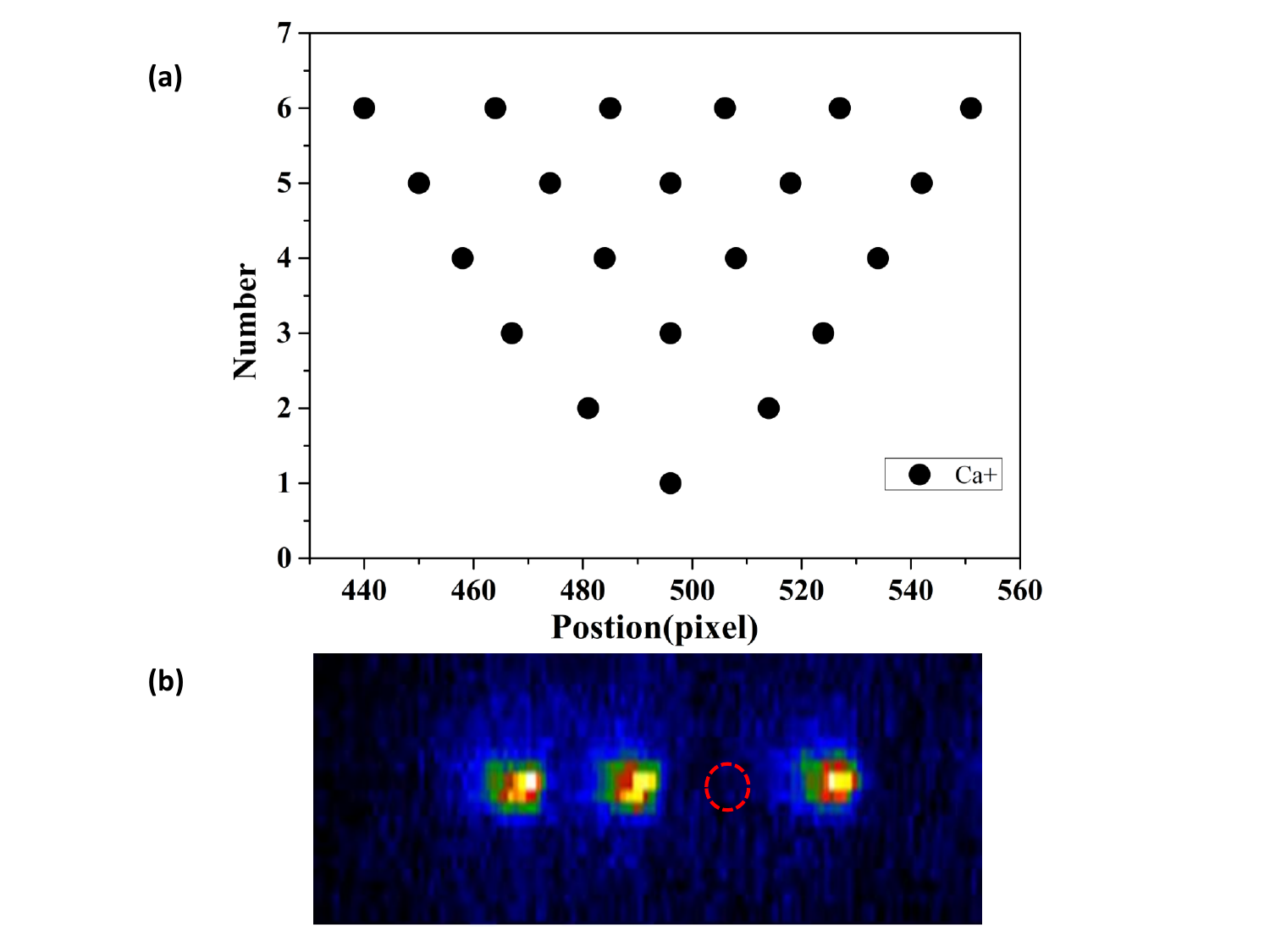}}
	\end{center}
	\caption{
	The relationship between ion distance and number of ions and images of trapped ions on EMCCD. (a) The ion distance will increase with the decreasing number of ions. (b) The dotted line shows a dark ion, which may be $^{27}$Al$^+$ and the others are $^{40}$Ca$^+$ ions.
	}
\end{figure}
Loading a single dark ion, the current applied to the oven and the 394 nm laser power must be optimized firstly. In our experiment, the current applied to Al oven is 4 A and the power of 394 nm laser is 500 $\upmu$W with beam waist of 100 $\upmu$m. Here $\omega_r$ and $\omega_z$ for a single $^{40}$Ca$^+$ ion are 700 kHz and 297 kHz, respectively. When the position of the $^{40}$Ca$^+$ ion imaged in the EMCCD moves suddenly, it means that one or more dark ions have been trapped successfully. Then the number of dark ion can be confirmed according to the ion distance.

The dark ion may not be $^{27}$Al$^+$ ion while AlH$^+$ or AlOH$^+$ ion \cite{25}. For these dark ions, it is impossible to identify the ion species only from the EMCCD images. However $^{27}$Al$^+$ ion can be distinguished from other dark ion species by precisely measuring the frequency of secular motion. For distinguishing the $^{27}$Al$^+$ ion from AlH$^+$ ion, the error of measured secular motion frequency should be less than 2 kHz.
\begin{figure}
	\begin{center}
		\resizebox{0.5\textwidth}{!}{\includegraphics{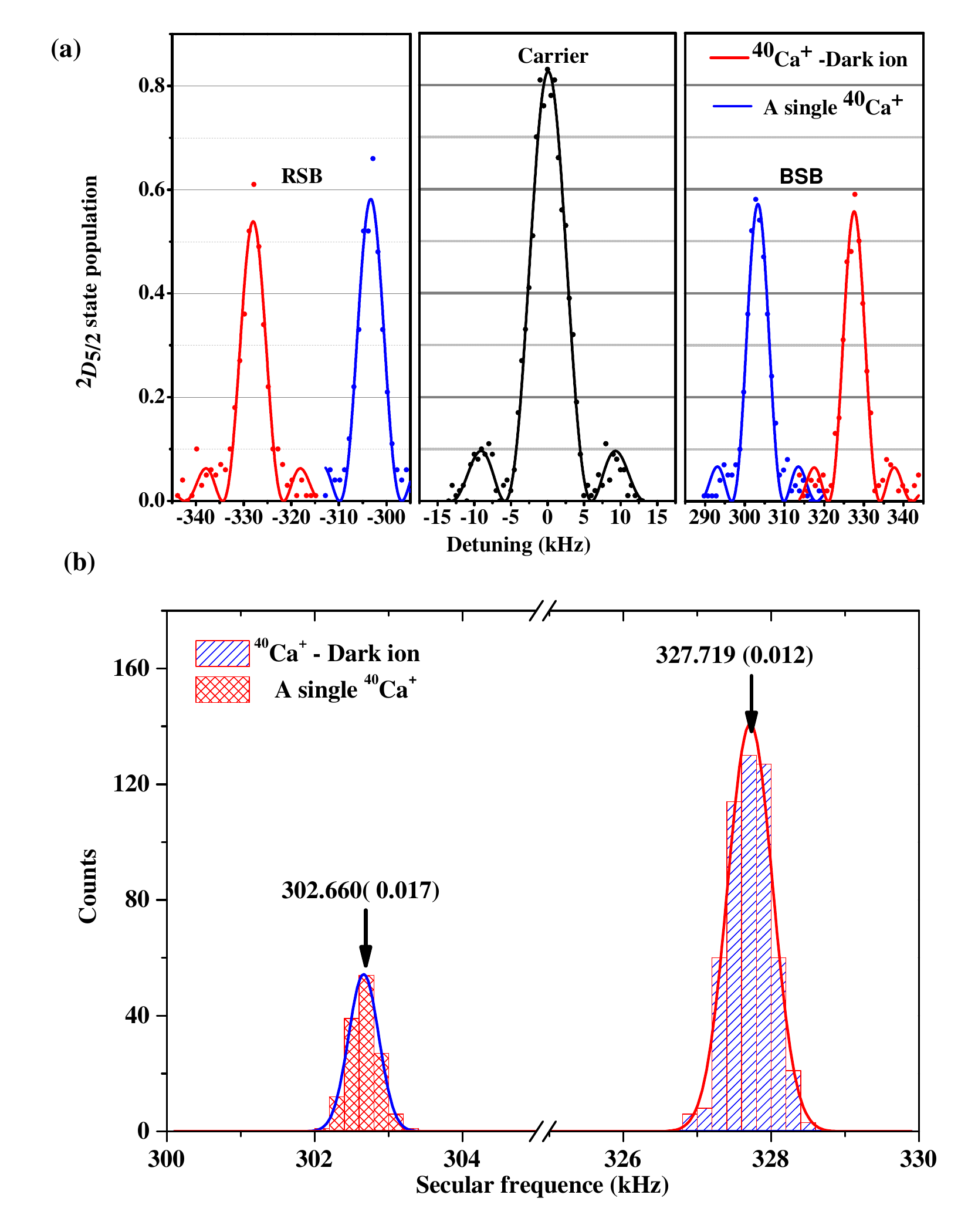}}
	\end{center}
	\caption{
 The axial sidebands ($\omega_z$) and carrier spectra of $^{40}$Ca$^+$ $^2S_{1/2} (1/2)-^2D_{5/2} (5/2)$ transition. (a) The black line is the carrier transition. The blue and red lines are the axial sidebands of $^{40}$Ca$^+$ and $^{40}$Ca$^+$-$^{27}$Al$^+$  ion pair crystal, respectively. (b) The statistics of axial blue sideband frequency for a single $^{40}$Ca$^+$ ($302.660(0.017)$ kHz) and $^{40}$Ca$^+-^{27}$Al$^+$ ion pair crystal ($327.719(0.012)$ kHz). The dark ion mass is calculated to be $26.98(0.01)$ a.u..
 }
\end{figure}
The secular motion frequency can be measured by interrogating the red and blue sideband spectra of $^{40}$Ca$^+$ ion $^2S_{1/2} (-1/2)-^2D_{5/2} (-5/2)$ transition with a probe laser at 729 nm. This laser is locked to a high-fineness ultra-low-expansion (ULE) cavity using Pound-Drever-Hall (PDH) technique. The linewidth of 729 nm laser is about 20 Hz, the precision of the secular sideband frequency is about 3 kHz resolution limiting by fluctuations of electric field and magnetic field. Using the four points locking scheme\cite{22}, the precision is improved by interrogating the RSB and BSB alternately. Fig.3 displays the measurement results under the condition of 400 V endcap voltage with 80 $\upmu$s probe time of the 729 nm laser. Fig.3 (a) is the single measurement of axial secular motion frequency. The black line is the carrier spectrum of single $^{40}$Ca$^+$. Blue lines are the $z$-axial secular motion frequency of single $^{40}$Ca$^+$ and red lines are the $z$ axial COM spectra $^{40}$Ca$^+$-$^{27}$Al$^+$ of ion pair crystal. Axial sideband spectra of a single $^{40}$Ca$^+$ ion and of $^{40}$Ca$^+$-dark ion pair crystal have been measured for 140 and 500 times respectively (Fig.3 (b)). The average axial secular frequency and COM secular frequency for single $^{40}$Ca$^+$ ion and $^{40}$Ca$^+$-dark ion pair crystal is $302.660 (0.017)$ kHz and $327.719 (0.012)$ kHz respectively. According to Eq.(6), the mass of the dark ion is calculated to be $26.98 (0.01)$ a.u., which testifies the dark ion to be an $^{27}$Al$^+$ ion.

The temperature of $^{40}$Ca$^+-^{27}$Al$^+$ ion pair crystal along axial COM has been estimated by comparing the relative sideband intensities \cite{26,27}. The probe laser linewidth of 729 nm is broadened by modulating the frequency of AOM in order to reduce the number of data points required for the sweep. The intensities of carrier and sideband spectra are measured to be 0.44 and 0.22 with 10 ms interrogation time. According to Ref. \cite{27}, the temperature of $^{40}$Ca$^+-^{27}$Al$^+$ ion pair crystal along axial COM is calculated to be about 7 mK.

In summary, in a linear Paul trap, a dark ion has been obtained by sympathetic cooling with a single $^{40}$Ca$^+$ ion. The mass of dark ion is measured by secular motion sideband spectra and it is proved to be $^{27}$Al$^+$ ion. The temperature of $^{40}$Ca$^+-^{27}$Al$^+$ ion pair crystal is estimated to be about 7 mK. The synthesis of $^{40}$Ca$^+-^{27}$Al$^+$ ion pair crystal is a milestone on progress of $^{27}$Al$^+$ ion quantum logic optical clock in the future.

We thank professor P.O.Schimidt, Dr. Yao Huang and Dr.Ting Chen for fruitful discuss. This work is supported by the National High Technology Research and Development Program of China (863 Program, Grant No. 2012AA120701) and the National Natural Science Foundation of China (Grant No. 11174326).

\end{document}